\documentclass[12pt]{iopart}
\input{psfig.sty}
\newcommand{\ar}{\arrowvert}
\newcommand{\ra}{\rangle}
\newcommand{\la}{\langle}

\newcommand{\cd}{\! \cdot \!}
\newcommand{\be}{\begin{equation}}
\newcommand{\ee}{\end{equation}}
\newcommand{\ba}{\begin{eqnarray}}
\newcommand{\ea}{\end{eqnarray}}

\begin{document}

\title[ Bulk viscosity in the cold CFL superfluid]
{Bulk viscosity in a cold CFL superfluid}

\author{Cristina Manuel$^1$ and Felipe Llanes-Estrada$^2$}
\address{$^1$ Instituto de Ciencias del Espacio (IEEC/CSIC) \\
Campus Universitat Aut\` onoma de Barcelona,
Facultat de Ci\` encies, Torre C5 \\
E-08193 Bellaterra (Barcelona), Spain \\
$^2$  Departamento de F\'{\i}sica Te\'orica I,  Universidad
Complutense, 28040 Madrid, Spain}

\ead{$^1$cmanuel@ieec.uab.es, $^2$fllanes@fis.ucm.es}

\date{today}

\begin{abstract}
We compute one of the bulk viscosity coefficients of cold CFL quark matter in
the temperature regime where the contribution of mesons, quarks
and gluons to transport phenomena is Boltzmann suppressed. In that
regime dissipation occurs due to  collisions of superfluid
phonons, the Goldstone modes associated to the spontaneous
breaking of baryon symmetry. We first review the hydrodynamics of
relativistic superfluids, and remind that there are at least three
bulk viscosity coefficients in these systems. We then compute the
 bulk viscosity coefficient associated to the normal fluid
component of the superfluid. In our analysis  we use Son's
effective field theory for the superfluid phonon, amended to
include scale breaking effects proportional to the square of the
strange quark mass $m_s$. We compute the bulk viscosity at leading order
in the scale breaking parameter, and find that it is dominated by
collinear splitting and joining processes.  The resulting transport coefficient
 is $\zeta=0.011\ m_s^4/T$, growing  at low temperature $T$
until the phonon fluid description stops making sense.
Our results are relevant to study
the rotational properties of a compact star formed by CFL quark
matter.
\end{abstract}

\maketitle

\section{Introduction}

In this paper we present the computation of one of the bulk viscosity coefficient
of color-flavor locked (CFL) quark matter \cite{Alford:1999mk} at low temperature.
This work represents a follow up of Ref. \cite{Manuel:2004iv}, where the shear
viscosity in the CFL phase was computed in the cold regime where the contribution
of all the gapped particles (mesons, quarks and gluons) is Boltzmann suppressed.

The density reached in the core of neutrons stars might be so high that all the hadrons
could be melted into their fundamental constituents. This consideration has motivated the
studies of QCD at
high baryonic density and low temperature \cite{reviews}. At least in the asymptotic high density regime,
when the QCD coupling constant is small, reliable theoretical predictions
for the behavior of quark matter can be formulated. Further, it has been known for long time that
cold dense quark matter should exhibit the phenomenon of color superconductivity.
In order to connect  theoretical predictions  with  possible astrophysical
signatures of quark matter \cite{Weber:2004kj}, it is important to have a precise knowledge of both the equation of state of quark matter and also of all transport coefficients, which are very sensitive to the presence of superconductivity
in the system.

It has been established that the viscosities put stringent tests
to astrophysical models for very rapidly rotating stars, such as
for millisecond pulsars. This is based on the existence of
r(otational)-mode instabilities in all relativistic rotating stars
\cite{Andersson:1997xt}, which are only suppressed by sufficiently
large viscosities. So the viscosities allows to discard
unrealistic models for millisecond pulsars.  There are many
different color superconducting phases, and their occurrence
depends on both the values of the baryonic chemical potential and
of the different quark masses. At present, there are several
computations of viscosity coefficients in the different quark
matter phases
\cite{Sawyer:1989uy,Madsen:sx,Madsen:1999ci,Sa'd:2006qv,Alford:2006gy,Alford:2007rw,Dong:2007mb}.
All of them have the motivation of studying the development of the
r-modes of a compact star made of, partly or entirely,  unconfined
quark matter.

Here we will only be concerned about the CFL phase, which is the
preferred phase in the presence of three light quark flavors. The
CFL  phase is special in many ways, as its long distance physics
is very similar to the corresponding one of QCD in vacuo
	\cite{Alford:1999mk,Schafer:1998ef}. Here we want to stress that
it is also very peculiar for its hydrodynamics. In the CFL case
the baryon symmetry is spontaneously broken, and thus CFL quark
matter becomes a superfluid of the same sort as those found in
condensed matter systems, such as in Bose-Einstein condensates.
Landau developed the hydrodynamical description of these
non-relativistic superfluids, proposing his famous two-fluid model
\cite{landaufluids,IntroSupe}. The hydrodynamics associated to
relativistic superfluids has been much less studied, although the
two-fluid model has been generalized to the relativistic domain
\cite{khalatnikov,Carter-Kha,Carter-Lang,Son:2000ht}. We believe
that the CFL superfluid may represent one specific and beautiful
example where the sophisticated relativistic superfluid
hydrodynamics could be derived from first principles, at least in
the asymptotic high density domain. One of the peculiarities of
these superfluids is that they have  more viscosities than a
normal fluid, as one can define more than one hydrodynamical
velocity.

In this article we compute the  bulk viscosity coefficient associated to
the normal fluid component of the cold CFL superfluid. Transport coefficients
are very sensitive to the temperature $T$ of the system. In the regime where $T$ is
smaller than all the energy gaps of all the quasiparticles (mesons, quarks and gluons),
transport coefficients in the CFL phase are dominated by the collisions of the
superfluid phonons \cite{Shovkovy:2002kv,Manuel:2004iv}, and these will be the
relevant modes in our study. In our analysis  we use Son's
effective field theory for the superfluid phonon \cite{Son:2002zn}. However, Son's theory
is scale invariant, and leads to a vanishing bulk viscosity, as this transport coefficient
measures the dissipation after a volume compression or expansion of the system.
 We then consider scale breaking effects due to a non-vanishing value of 
the strange quark mass. The inclusion of quark masses, that
represent an explicit chiral symmetry breaking effect in the QCD
Lagrangian,  makes the octet of (pseudo) Goldstone bosons of the
CFL phase (the pions, the kaons, the eta) massive. Because a quark
mass term in the Lagrangian respects the baryon symmetry, the
superfluid phonon remains massless, although quark mass effects
still affect its dynamics, as we will see. We compute the bulk
viscosity at leading order in this scale breaking parameter, and
find that it is dominated by collinear splitting and joining
processes. Surprisingly, the computation shares many technical
similarities with that of the bulk viscosity in the hot, weakly
coupled, phase of QCD at zero chemical potential
\cite{Arnoldbulk}, as we will later point out.

Let us stress that at higher temperatures other quasiparticle  modes might be relevant
for bulk viscosity as well.
In Ref.~\cite{Alford:2007rw} the bulk viscosity due to kaons in the CFL phase has been computed. Allowing
for flavor changing processes, mediated by the electroweak interactions, the bulk viscosity
has been computed assuming that the relevant processes are those of a neutral kaon decaying into two superfluid phonons, and to $K^\pm \leftrightarrow e^\pm + \nu$. However, as found in that reference, for temperatures
below the energy gap  associated to the kaon, $\delta m$, the kaon contribution to bulk
viscosity is exponentially suppressed $\sim e^{-\delta m/T}$,  as naturally expected. There is an additional uncertainty
of at what temperatures this suppression is effective, as the value of the kaon masses
computed in the literature can only be trusted in the asympotic high density limit.
They are believed to be in the range of the few MeV, or slightly higher
 \cite{Son:1999cm,Bedaque:2001je,Manuel:2000wm,Schafer:2001za,Schafer:2002ty,Ruggieri:2007pi}.

This paper is structured as follows. In Sec.~\ref{Sec-hydro} we recall the hydrodynamical equations
of a relativistic superfluid. Sec.~\ref{Sec-sonic} is devoted
to review Son's effective field theory for the superfluid phonon. With the same effective field theory,
one can easily also derive the dispersion law for the phonon in a moving superfluid, introducing the
concept of acoustic or sonic metric, which also allows us to identify
these phonons with the sound waves of the superfluid. In
Sec.~\ref{Sec-scalebreak} we see how  scale breaking effects could be
included in Son's Lagrangian. The explicit computation of the bulk
viscosity is given in Sec.~\ref{Sec-scalebreak}. After identifying the
leading collisional processes relevant for this transport coefficient,
we write down the Boltzmann equation for the phonon, and linearize it
in the small deviations around equilibrium in Subsec.~\ref{Sec-bulk}.
The collision term is explicitly written down
in Subsec.~\ref{Sec-coll}, and the numerical results of our computation are displayed in Subsec.~\ref{Sec-num}.
We conclude with a discussion of our results in Sec.~\ref{Sec-disc}. We will use throughout
natural units  $\hbar = c = k_B = 1$ and the metric conventions $(1,-1,-1,-1)$.

\section{Hydrodynamics in relativistic superfluids}
\label{Sec-hydro}


In this Section we present a quick review of the hydrodynamical equations for a relativistic superfluid.
These represent the natural relativistic generalization of Landau's two fluid model of  superfluid (non-relativistic) dynamics \cite{landaufluids,IntroSupe}.
There are different formulations of the hydrodynamics of a relativistic superfluid
\cite{khalatnikov,Carter-Kha,Carter-Lang,Son:2000ht}, they all differ in the choice
of the hydrodynamical variables used to describe the fluids.

The hydrodynamical equations in the superfluid take the form of
conservation laws, as in
a normal fluid. If $n^\rho$ is
the particle current (in our case, the baryon current) these are
\begin{equation}
\label{sup-hydroeqs1}
\partial_\rho n^\rho   = 0 \ ,
\end{equation}
and the energy-momentum conservation law
\begin{equation}
\partial_\rho T^{\rho \sigma} = 0 \ .
\end{equation}

In the absence of dissipation, the entropy current is conserved, and thus one further
has
\begin{equation}
\partial_\rho s^\rho   =   0  \ .
\end{equation}

In the superfluid,  the  gradient of the phase of the condensate allows to define the
four vector
\begin{equation}
\label{sup-covect}
\mu_\rho = \partial_\rho \varphi \ .
\end{equation}

The approach of Carter and Khalatnikov \cite{Carter-Kha}
defines all the hydrodynamical equations based on
expressing both the particle current and the energy-momentum tensor in terms of
$\mu_\rho$ (or $\varphi$) and $s_\rho$. In particular, in Ref.~\cite{Carter-Lang} it is shown that the
energy-momentum tensor can be written as
\begin{equation}
\label{Sup-enermom}
T^{\rho \sigma} = A \mu^\rho \mu^\sigma + B s^\rho s^\sigma - P g^{\rho \sigma} \ ,
\end{equation}
where $P$ is the generalized pressure of the system. The coefficients $A$ and $B$  can be obtained
with the knowledge of $P$. Similarly, $n^\rho$ can be expressed in terms of both $\mu^\rho$ and $s^\rho$.

Son formulated a different description of the hydrodynamics of the
relativistic superfluids in Ref.~\cite{Son:2000ht}.  After a non-trivial
mapping of his variables, their equations can be converted to the Carter
and Khalatnikov form \cite{private}.

 In the zero temperature limit, when the pressure is only a function of
the chemical potential,  the entropy current vanishes. It is only  in  this case when the
the energy-momentum tensor takes the form of that of an ideal fluid \cite{Carter-Lang}.
If we define the velocity vector
\begin{equation}
u^\rho = \frac{\mu^\rho}{\mu}
\end{equation}
such that is properly normalized, $u^\rho u_\rho = 1$, then in the
cold $T \to 0$ limit
one has \cite{Carter-Lang}
\begin{equation}
T^{\rho \sigma} = ( n \mu) u^\rho  u^\sigma - P g^{\rho \sigma} =  \left(\epsilon+P\right) u^\rho  u^\sigma - P g^{\rho \sigma}
\end{equation}
where $\epsilon$ is the energy density of the system, and we have used the zero temperature relation
 $n \mu = \epsilon+P$. In this case the associated hydrodynamical equations are the same as in an ideal fluid.
The entropy strictly vanishes, and thus, there is no dissipation.

At finite temperature, the entropy does
not vanish, and dissipational processes are responsible for entropy production.
Dissipative relativistic superfluid hydrodynamical equations  have only been derived, to the best of
our knowledge, in Ref.~\cite{khalatnikov}, although there is a vast literature on the subject for
non-relativitistic superfluids \cite{landaufluids,IntroSupe}. In Ref.~\cite{khalatnikov}, and
after imposing that deviations from the dissipationless
particle current and energy-momentum are expressed in terms of linear gradients of the basic hydrodynamical
variables, and  that entropy production is positive-definite, it was found that
more kinetic coefficients that in a normal fluid can be defined.
We leave for a future project a much more detailed discussion on all the possible
transport coefficients that can be defined in the CFL superfluid.  We
simply note here that in
a non-relativistic superfluid four viscosity coefficients can be
defined \cite{landaufluids,IntroSupe}, and thus, at a minimum, the
same number of viscosity coefficients
in the relativistic domain exist.

In this paper we compute the  bulk viscosity coefficient
associated to the normal fluid.  For that purpose, and taking into
account that the bulk viscosity is a Lorentz
scalar,  we will work in the superfluid rest frame, as this simplifies
enormously the computation. More specifically, we will compute
the dissipative term in the energy-momentum tensor that goes as
\begin{equation}
T^{i j}_d = - \zeta \,\delta^{ij}\,{\bf \nabla} \cd {\bf V}
\end{equation}
where ${\bf V}$
 is the  velocity of the normal fluid in the
superfluid rest frame.

\section{The cold CFL superfluid and the sonic metric}
\label{Sec-sonic}

In this Section we review the effective field theory of the superfluid
phonon constructed by Son \cite{Son:2002zn}. Further, we introduce the
concept of sonic or acoustic metric, which is rather convenient in order
to describe the dynamics of the phonon moving in the background of the
superfluid. It also allows us to give the interpretation of the
superfluid phonon as a sound wave.

In a non-relativistic superfluid, gravity analogues for the
description of the low energy collective modes or superfluid
phonons have been developed (see
\cite{Barcelo:2005fc,Volovik:2000ua} and references therein). In such an
approach, one treats the superfluid  as a gravitational
background, in which the quasiparticles, composing the normal
fluid, propagate. We will use the same analogy here.

Son showed that the effective field theory  for the only truly
Goldstone boson of the CFL phase can be constructed from the equation of
state (EOS) of normal quark matter \cite{Son:2002zn}. If $\varphi$ is
the phase of the condensate, and one defines
$D_\rho  \varphi \equiv \partial_\rho \varphi - (\mu,{\bf 0})$,
then the effective Lagrangian for $\varphi$ is expressed as
\begin{equation}
{\cal L}_{\rm eff}[D_\rho \varphi] = P [\mu = (D_\rho \varphi D^\rho  \varphi)^{1/2}] \ ,
\end{equation}
where $P$ is the pressure of the system at zero temperature.

At asymptotic large densities the EOS of CFL quark matter  reads
\begin{equation}
 \label{freeEOS}
P [\mu]
=\frac{3}{4\pi^2} \mu^4 \ ,
\end{equation}
where $\mu$ is the quark chemical potential. At very high $\mu$,
when the coupling constant is small $g (\mu) \ll 1$, the effects
of interactions and the effects of Cooper pairing are subleading
and neglected in Eq.~(\ref{freeEOS}). Also, one assumes that the
quark masses give a subleading effect, as $m_q \ll \mu$. From Eq.
(\ref{freeEOS}) Son obtained the effective Lagrangian
\begin{equation}
\label{L-BGB-0}
{\cal L}_{\rm eff}  = \frac{3}{4 \pi^2}
\left[ (\partial_0 \varphi - \mu)^2 - (\partial_i \varphi)^2
\right]^2 \ .
 \end{equation}

There is an interesting  interpretation of the equations of motion
associated to $\varphi$. The classical equations of motion derived from
the above Lagrangian can be re-written as the hydrodynamical
conservation law of a current representing baryon number flow,
\begin{equation}
\label{S-hy-1}
 \partial_\nu (n_0 \bar u^\nu) = 0 \ ,
\end{equation}
 where $n_0 =\frac{dP}{d \mu} |_{\mu =\mu_0}$ is interpreted
as the baryon density \cite{Son:2002zn}. Son defined
the superfluid velocity $\bar u^\rho$ as being proportional to the gradient of the
condensate phase
\cite{landaufluids},
\begin{equation} \label{svelocity}
\bar u_\rho = - \frac{D_\rho \bar \varphi}{\mu_0} \ ,
\end{equation}
where  $\mu_0 = (D_\rho \bar \varphi D^\rho \bar \varphi)^{1/2}$. It only differs from the
choice of the last Section by an irrelevant constant, although the hydrodynamics is completely
analogous.
The  energy-momentum tensor associated to the theory described in Eq.~(\ref{L-BGB-0})
 can also be written in terms of the velocity
defined in Eq.~(\ref{svelocity}) and Noether's energy-density $\epsilon$,
 \begin{equation}
\label{S-hy-2}
T^{\rho \sigma} = (\epsilon + P) u^\rho u^\sigma - g^{\rho \sigma} P \ .
 \end{equation}
It is conserved and traceless
\begin{equation} \label{Tconserved}
\partial_\rho T^{\rho \sigma} = 0\ ,\qquad  T^\rho_\rho=0 \ .
\end{equation}

Eqs.~(\ref{S-hy-1}) and (\ref{Tconserved}) are the hydrodynamic
equations for the relativistic superfluid  \cite{Son:2002zn}.
They need modifications at  finite temperature as explained in the
previous Section. At low temperatures, the superfluid phonons
are thermally excited and conform the normal fluid component, which are
responsible for the entropy current in the system. Other particles may
also conform an additional component to the normal fluid, but at low temperatures,
as discussed in the Introduction, their contribution to the hydrodynamics
is Boltzmann suppressed.

Let us also  mention that Son's Lagrangian yields the effective field
theory of the phonons moving in the background of the superfluid.
The phonon is a Goldstone boson, given also by the phase of the
condensate. When the superfluid is at rest, their interactions are
given in Eq.~(\ref{L-BGB-0}). To find the phonon dispersion relation in
a moving superfluid we will simply consider the quantum fluctuations
around the classical solution of
Eq.~(\ref{L-BGB-0})
\begin{equation}
\varphi (x) = \bar \varphi(x) + \phi(x) \ .
\end{equation}

The action associated to Son's Lagrangian
\begin{equation}
S[\varphi] = \int d^4 x \, {\cal L}_{\rm eff}[\partial \varphi]
\end{equation}
is then expanded around the classical solution
\begin{equation}
S[\varphi] = S[\bar \varphi] + \frac 12 \int d^4 x \,\left \{ \frac{
\delta^2 {\cal L}_{\rm eff} }
{\delta(\partial_\mu \varphi) \delta(\partial_\nu \varphi)}
\right \} \partial_\mu \,\phi \partial_\nu \phi + \cdots
\end{equation}

The equation of motion of the linearized fluctuation - here the
superfluid phonon - can be written as that of a boson moving in a
non-trivial gravity background
\begin{equation}
\partial_\mu \left( \sqrt{-  {\cal G}} \,  {\cal G}^{\mu \nu}
\partial_\nu \phi \right) = 0
\end{equation}
where in this case, we identified
\begin{equation}
\sqrt{- {\cal G} } \, {\cal G}^{\mu \nu} =  \frac{\delta^2{\cal L}_{\rm
eff}}{\delta(\partial_\mu \varphi)
\delta(\partial_\nu \varphi)} \Bigg \vert_{\bar \varphi} = \frac{3
\mu^2}{2 \pi^2} \left \{ g^{\mu \nu} + \left(\frac {1}{c_s^2} - 1
\right) \bar u^\mu \bar u^\nu \right \}
\end{equation}
where $c_s = \frac{1}{\sqrt{3}}$ is the speed of sound in the ultrarelativistic system.
Thus, from Son's Lagrangian, we have derived the so-called sonic or
acoustic metric tensor
${\cal G}^{\mu \nu}$ \cite{Carter-Lang,Barcelo:2005fc}.
One then finds the phonon dispersion relation in the moving superfluid
as solutions of the equation
\begin{equation}
\label{disp-eq}
{\cal G}^{\mu \nu} k_\mu k_\nu = 0 \ ,
\end{equation}
where $k_\mu = (E, {\bf k})$.

Let us note that in an ideal fluid the fluctuations of the pressure,
$\delta p$, obey  the sound wave equation (see e.g., Eq.~(23) of
Ref.~\cite{Mannarelli:2007gi}), which in Fourier space reads,
\begin{equation}
\label{sound-wave}
\frac{1}{\bar u^\mu k_\mu} \left\{ \left(\frac{1}{c_s^2} -1 \right) (\bar u^\mu k_\mu)^2 + k^2    \right\}
\delta p = 0 \ .
\end{equation}
Simply by defining
\begin{equation}
\delta p = \frac{3 \mu^2}{2 \pi^2} (\bar u^\mu \partial_\mu)\, \phi
\end{equation}
we derive the same dispersion equation (\ref{disp-eq}). This is
naturally so, as the superfluid phonons
describe the sound waves associated to the superfluid component of the fluid.
Let us stress here that in the relativistic superfluid there are two (first and second) sound speeds,
associated to the fact that there are two different fluid components \cite{Carter-Lang}.

In the superfluid rest frame, that is, where $\bar u^\mu = (1,0,0,0)$ the phonon dispersion relation simplifies to the
form
\begin{equation}
E_k = c_s k \ ,
\end{equation}
 as it can be easily checked. This is the frame where we will perform our computations.
However, we want to stress the fact that this would not be correct in order to compute other
transport coefficients in the system that involve the superfluid velocity $\bar u^\mu$. In those cases,
the gravity analogues described here become a very efficient tool, which will be exploited in future references.
Although  sonic metrics are usually employed to construct simple
laboratory analogue models of Einstenian gravity \cite{Garay:2000jj},
we think the method also shows promise in reverse, for  the involved
features of relativistic superfluids might be conceptually understood
in the framework of gravity analogues.

\section{Effective field theory for the superfluid phonons including
scale breaking effects}
\label{Sec-scalebreak}

In Ref. \cite{Manuel:2004iv} the effective field theory Lagrangian (\ref{L-BGB-0}) was used
to compute the shear viscosity in the phonon fluid of cold CFL quark matter. For the computation
of the bulk viscosity  one has to introduce corrections to that effective field theory.
As signaled by the vanishing of the trace of the energy-momentum tensor, Eq.~(\ref{Tconserved}),
this is an scale invariant theory. Bulk viscosity is a transport coefficient that measures
the dissipation after a volume compression or expansion, and it vanishes exactly for a
scale invariant relativistic theory. Thus, in this Section we will look for corrections to Eq.~(\ref{L-BGB-0})
that introduce scale breaking effects.

The quantum scale anomaly breaks the conformal symmetry in
the system introducing, through dimensional transmutation, the quantum scale $\Lambda_{\rm QCD}$.
One could then compute $g$-corrections to Son's effective field theory, arising as
$g$-corrections to the pressure of quark matter. These corrections would introduce new terms
in the superfluid phonon Lagrangian proportional to the QCD beta function.
In the very high $\mu$ limit, when $g (\mu) \ll 1$, we expect this to be a rather negligible
effect.

The inclusion of quark mass effects in the system also breaks scale
invariance. In  CFL quark
matter, the quark masses represent an explicit chiral symmetry breaking effect, which gives masses
to the associated (pseudo) Goldstone bosons (the pions, kaons and eta). A quark mass term
in the QCD Lagrangian respects baryon symmetry, and thus it does not make the superfluid phonon massive,
however it still affects its dynamics, as we show below.

Since all three light quarks participate in the CFL phase,
the largest effect comes from the  strange quark mass $m_s$,
and we ignore que masses of the
up and down quarks, as $m_u, m_d \ll m_s$. Further, we will consider that
$m_s^2 < 2 \Delta \mu$
\cite{reviews},
which is the threshold value under which the CFL phase is stable.
We will always work in a leading order expansion in $m_s^2/\mu^2$.
After imposing the constraints of electrical neutrality and beta equilibrium of quark matter,
the first correction at order $m_s^2/\mu^2$ to the pressure reads \cite{Alford:2002kj}
\begin{equation}
P [\mu]
=\frac{3}{4\pi^2} \left(\mu^4 - \mu^2 m^2_s \right) \ .
\end{equation}
To this order, first in the $m_s^2/\mu^2$ expansion, and since isospin
breaking effects happen to be of the same order, one must specify that
$\mu$ refers to precisely $\mu_n/3$ in terms of the baryon chemical
potential. Let us stress here that as in Son's theory,
 Eq.~(\ref{freeEOS}), we neglect both the effects of interactions and
of Cooper pairing, assuming the asymptotic high density, and thus the
weak coupling domain. The expansion here is aimed to consider the
most relevant scale breaking effect that corrects Son's Lagrangian.

Following the same procedure advocated by Son, with the knowledge of
the pressure we get the effective field theory for the phonons or sound
waves in the system
\begin{equation}
\label{L-BGB-mass}
{\cal L}_{\rm eff}^m  = \frac{3}{4 \pi^2}
\left[ \left((\partial_0 \varphi - \mu)^2 - (\partial_i \varphi)^2 \right)^2
- m^2_s \left((\partial_0 \varphi - \mu)^2 - (\partial_i \varphi)^2 \right)
\right] .
 \end{equation}

We now rescale the phonon field
\begin{equation}
{\tilde \varphi} = \frac{3 \mu}{\pi} \sqrt{1 - \frac{m^2_s}{6 \mu^2}} \, \varphi
\end{equation}
to normalize the kinetic term in accordance with the LSZ formula.
Then the Lagrangian for the rescaled field reads
\begin{equation}
 \label{L-BGB}
 {\cal L}_{\rm eff}  = \frac 12 (\partial_0 {\tilde \varphi})^2 -
\frac{c_s^2}{2} (\partial_i {\tilde \varphi})^2 - g_3
\partial_0 {\tilde \varphi} (\partial_\mu {\tilde \varphi}  \partial^\mu {\tilde \varphi}) +
g_4(\partial_\mu {\tilde \varphi}  \partial^\mu
{\tilde \varphi})^4 \ ,
\end{equation}
where, to order $m_s^2/\mu^2$ we find
\begin{eqnarray}
\label{m-speedsound}
c_s^2 & = & \frac 13 \left( 1 - \frac{m^2_s}{3 \mu^2} \right) \ , \\
g_3 & = & \frac {\pi}{9 \mu^2} \left( 1 + \frac{m^2_s}{4 \mu^2} \right) \ , \qquad
g_4  =  \frac {\pi^4}{108 \mu^4} \left( 1 + \frac{m^2_s}{3 \mu^2} \right) \ .
\end{eqnarray}
Thus, the leading order effect of the strange quark mass in the phonon
Lagrangian is to modify the velocity of the phonon, that obviously equals to the speed of
sound in the conformally broken system,  and a finite renormalization of the cubic and
quartic self-couplings.

\section{Transport theory in the phonon fluid}

\begin{figure}
\hspace{4cm}
\psfig{figure=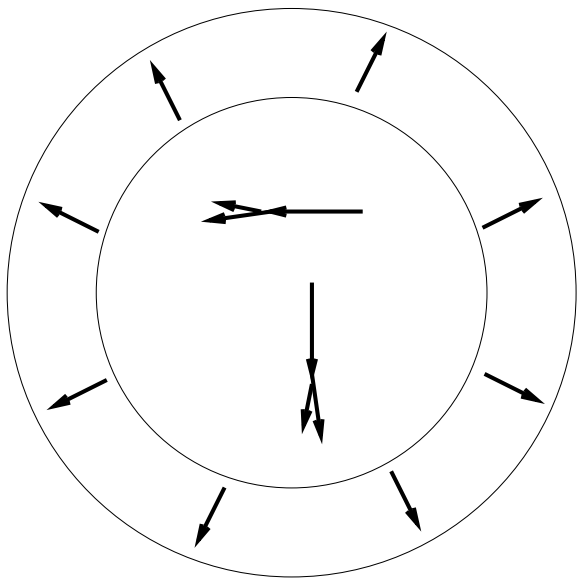,height=3.0in}
\caption{Under spherically symmetric, radially non-uniform rarefaction
(compression), of a gas not invariant under dilatations,  the pressure
diminishes below (increases above) its equilibrium value. For the
superfluid phonon gas, equilibrium is restored by phonon collinear
splitting (joining).}
\label{fig:expansion}
\end{figure}

We now turn to the microscopic description of the bulk viscosity at low temperature.
The dominant process relevant for the bulk viscosity, shown in
Fig.~\ref{fig:expansion}, is phonon collinear splitting or joining processes $1 \leftrightarrow 2$.  Large angle $2 \leftrightarrow 2$
scatterings are suppressed, as compared to collinear splitting, by powers of $1/\mu^2$
as found in Ref.\cite{Manuel:2004iv}. Because we are considering the regime $T \ll \mu$,
we can safely neglect those collisions. Further, we don't consider
small angle $2 \leftrightarrow 2$ collisions, which are collinearly enhanced \cite{Manuel:2004iv}.
Considering   simultaneously  the $1 \leftrightarrow 2$ processes and small angle  $2 \leftrightarrow 2$ collisions
would mean to incur in a wrong double-counting.

It is remarkable that the computation in the asymptotic large density and low temperature
CFL phase has many points in common with the same computation in the very hot, weakly coupled
phase of QCD at vanishing chemical potential. In the hot phase of QCD, bulk viscosity is
dominated by both effective collinear splitting processes $1 \leftrightarrow 2$, as well as by $2 \leftrightarrow 2$ collisions
\cite{Arnoldbulk}.
Fortunately, in the CFL phase the computation is simpler, as the last processes are certainly suppressed.
However, we will find convenient to follow the same technical strategy that in Ref.~\cite{Arnoldbulk},
and tackle different subtle points in the computation in the
 same way as in that reference, which we will closely follow.

There is a a different technical remark that we would like to point out here.
  Collinear splitting is closed by a convex
dispersion relation when one considers corrections to the phonon dispersion relation
at order $k^2/\Delta^2$ \cite{Zarembo:2000pj}, where $\Delta$ is the superconducting gap.
If one considers those corrections, one
should then study other number changing collisions, such as a $2 \leftrightarrow 3$
scattering for the computation of the bulk viscosity. In these $2  \leftrightarrow 3$ processes, one
may treat one of the particles as a spectator whose only
role is to restore energy-momentum conservation in the splitting vertex.
We will ignore this subtlety here and operate as if the dispersion
relation was exactly linear, kinematically allowing for collinear
splitting and rejoining. We expect that considering those processes would simply allow
us to get corrections of order $T^2/\Delta^2$ to our leading result.

\subsection{The Boltzmann equation and bulk viscosity}
\label{Sec-bulk}

In the superfluid rest frame the out of equilibrium phonon distribution function
evolves according to the Boltzmann equation
\be
\frac{\partial f_p}{\partial t} + {\bf v}_{\bf p}\cd {\bf \nabla}_x f_p =
-C[f_p] \ ,
\ee
where ${\bf v}_{\bf p} = {\bf \nabla}_p E_p$, and $C[f]$ is the collision integral.
We have also introduced the short-hand notation $f_p = f (x, {\bf p})$ that we will
use in what follows.

We will consider small deviations from equilibrium
\be
f_p = f^{\rm eq}_p + f^1_p + \cdots \ .
\ee

In the phonon rest frame
\be
f^{\rm eq}_p=f^0_p=\frac{1}{e^{\beta
E_p}-1} \ ,
\ee
where $\beta = 1/T$.

The Boltzmann equation is linearized in the departures of equilibrium. One has to keep in mind
that the collision term evaluated with the equilibrium
function vanishes by detailed balance. As in Ref.~\cite{Arnoldbulk}, we cast
the left-hand-side of the Boltzmann
equation with separated variables as
\be
X(x) \beta f^0_p (1+f^0_p) q(p)
\ee
that parametrizes the advective derivative of $f$ by defining $q(p)$.
In order to do so, one needs some thermodynamical relations. In the superfluid
rest frame, those are $\epsilon + P = T \frac{dP}{dT}$, exactly as in the hot quark-gluon plasma
at zero chemical potential. The
explicit calculation of this advective term in \cite{Arnoldbulk} yields
\be
\label{adv-term}
q(p) =\frac{ {\bf p}\cd {\bf v}_p}{3} - c_s^2 \frac{\partial (\beta
E_p)}{\partial \beta} = \left( \frac 13 - c_s^2 \right) E_p \ ,
\ee
and for a departure of equilibrium that is a uniform compression,
\be
X(x)= {\bf \nabla} \cd {\bf V}
\ee
as relevant for the bulk viscosity.

Eq.~(\ref{adv-term}) reflects the fact that in a relativistic
scale invariant theory, where $c_s^2 = \frac 13$, $q(p)=0$ and thus the bulk viscosity
vanishes. Thus, we allow for scale breaking effects, as anticipated in
the previous Section. A non-vanishing strange quark mass,  $m_s\not = 0$,
introduces a correction to the speed of sound,  as found in Eq.~(\ref{m-speedsound}).
In this case we obtain
\be
q(p)=\frac{p^2}{9E_p} \frac{m_s^2}{\mu^2} \ ,
\ee
proportional to our scale  breaking parameter
$m_s^2/\mu^2$.

The next step in the analysis is to project the Boltzmann equation into
weak (integrated over $p$) form amenable to variational treatment.
To alight notation it is convenient to introduce a scalar product
\be
\la\  \ar \ \ra = \beta^3 \int \frac{d^3{\bf p}}{(2\pi )^3} \ .
\ee
Multiplying the Boltzmann equation by $-T^2$ we  obtain
\be \label{Boltzmanneq}
-Tf^0_p (1+f^0_p) \frac{p^2}{4E_p} \frac{m_s^2}{\mu^2} X(x)=
-T^2C[f^0_p+f^1_p] \ ,
\ee
and we define, as in \cite{Arnoldbulk}, a function for the left-hand-side
source $S$, without the $X(x)$ factor (that will cancel left and right
in the Boltzmann equation), nor the conformal symmetry breaking factor,
$(m_s/\mu)^2$,
\be \label{defofs}
S(p) = -Tf^0_p (1+f^0_p) \frac{p^2}{4E_p} \ .
\ee
The projection of this equation over a complete orthonormal basis of
functions $\psi_n$  is then, formally
\be
X(x) \frac{m_s^2}{\mu^2} \la \psi_n \ar S \ra = \la \psi_n \ar
C[f^0_p+f^1_p] \ra \ .
\ee
And we further introduce a dimensionless variable (henceforth, the
bar notation denotes an adimensional function of momenta over temperature)
$$
S(p)= -T^2 \bar{S}(p/T) \ .
$$
The bulk viscosity is given as the matrix element
\be \label{bulkvisco}
\zeta = \frac{m_s^4}{\mu^4} \la S \ar \left( \frac{\delta C}{\delta
f}\right)^{-1} \ar S \ra
\ee
with the differential matrix of the collision operator
\be
\left(  \frac{\delta C}{\delta f}\right)_{mn}=
\la \psi_m \ar  \frac{\delta C[f]}{\delta f} \ar \psi_n\ra \ .
\ee

\subsection{The collision term}
\label{Sec-coll}

Bulk viscosity involves the relaxation of a momentum gradient along the
same direction of the momentum. Within Son's effective theory, the lowest
order effect that achieves this is the collinear splitting induced by the
cubic term. Because our computation
is done at leading order in the scale breaking parameter, $m^2_s/\mu^2$,
and this has already taken into account in Eq.~(\ref{adv-term}), it is enough
to keep the scattering matrices as arising in the scale invariant theory.
Thus, in this Section, we take the value  $c_s=1/\sqrt{3}$.
Let us insist again on the fact that collisions involving two bosons in the initial state and two
bosons in the final state are suppressed by a further power of $\mu^2$ at
the amplitude level, and thus we neglect them.

We will need therefore the amplitudes for a boson of momentum $p$ to
split into two bosons of momenta $p'$, $k'$ and the amplitude for a boson
of momentum $p'$ to split and give back in the final state the $p$ boson
and a $k'$ boson. From Son's Lagrangian density these are found to be,
\begin{equation}
{\mathcal{M}}(p;p',k') =  \frac{-i2\pi}{9\mu^2} (p^0(p' \cdot k') +
p^{'0}(p \cdot k')+ k^{'0}(p \cdot p')) \ ,
\end{equation}
or,  employing momentum conservation and the linear dispersion relation
$p^0=c_s \ar p\ar$,
\be
\ar {\mathcal{M}}(p;p',k') \ar^2 = \frac{4\pi^2}{81\mu^4}
c_s^2 \ar p\ar^2 \ar k' \ar^2 4x^2(\ar p\ar -\ar k'\ar)^2
\ee
where $x=\hat{p}\cd \hat{k'}$.
Similarly, one finds
\be
\ar {\mathcal{M}}(p';p,k') \ar^2 = \frac{4\pi^2}{81\mu^4}
c_s^2 \ar p\ar^2 \ar k' \ar^2 4x^2(\ar p\ar +\ar k'\ar)^2 \ .
\ee

The form of the collinear splitting collision terms that enter into
the Boltzman equation can be read off reference \cite{AMY}
\ba \label{collisionterm}
C^{1\to 2}[f_p]= C^{1\to 2}_a[f_p] + C^{1\to 2}_b[f_p] \\
\nonumber
= \frac{1}{4E_p}[\int]_{p'k'}(2\pi)^4
\delta^{(4)}(p-p'-k')
\left( f_p (1+f_{p'})(1+f_{k'})-f_{p'}f_{k'}(1+f_p)\right)
\\ \label{collisionterm2}
+\frac{1}{2E_p} [\int]_{p'k'}(2\pi)^4\delta^{(4)}(p'-p-k')
\left( f_{p'} (1+f_p)(1+f_{k'})-f_pf_{k'}(1+f_{p'})\right) \ ,
\nonumber
\ea
where we introduced the shorthand notation
$$
[\int]_{p'k'} \equiv \int\int \frac{d^3{\bf k'}}{2E_{k'}(2\pi)^3}
\frac{d^3 {\bf p'}}{2E_{p'}(2\pi)^3}\ar {\mathcal{M}}(p;p',k') \ar^2 \ .
$$
Let us reduce Eq.~(\ref{collisionterm}) to a more tractable form. The idea
is to split the momentum conservation delta into an energy part, a
longitudinal part (defined along $p$), and a transverse part. The best way
to organize the calculation is to introduce a collinear splitting function

\ba \label{splitting}
\gamma(p;k',p') = \int \frac{d^2k'_\perp d^2p'_\perp}{(2\pi)^34 c_s^3\ar
p'\ar\ar k' \ar} \frac{\ar p\ar}{2} \\ \nonumber \ar
{\mathcal{M}}(p;p',k') \ar^2
\delta^{(2)}(p_\perp-p'_\perp-k'_\perp) \delta(E_p-E_{p'}-E_{k'}) \ ,
\ea
so that the first collision term in Eq.~(\ref{collisionterm})
becomes
\ba
C^{1\to 2}_a [f_p] =\frac{2\pi}{2\ar p \ar^2}\int_0^\infty \int_0^\infty
d\ar k'\ar d\ar p'\ar \gamma (p;p',k') \delta(\ar p\ar - p'_L -k'_L)
\\ \nonumber
\left( f_p (1+f_{p'})(1+f_{k'})-f_{p'}f_{k'}(1+f_p)\right)
\ea
Now, energy conservation in Eq.~(\ref{splitting}) forces the
transverse momentum to vanish
\be
\delta(c_s(\ar p\ar -\ar p'\ar -\ar k'\ar)) = \frac{\delta(\ar
k'_\perp\ar)}{c_s \left( \frac{k'_\perp}{\sqrt{p^{'2}_L+k^{'2}_\perp}}
+  \frac{k'_\perp}{\sqrt{k^{'2}_L+k^{'2}_\perp}}
\right)} \ .
\ee
Thus
\be
\gamma(p;p'=p-k',k')=\frac{1}{(2\pi)^28 c_s^4} \left( \frac{16\pi^2
c_s^2}{81\mu^4} \right) \ar p\ar^2 \ar k'\ar^2(\ar p \ar - \ar k'
\ar)^2 \ .
\ee

Finally one can express the collision term as
\ba
\label{coll-term}
C^{1\to 2}_a=\frac{2\pi}{2\ar p \ar^2}\int_0^\infty \int_0^\infty
d\ar k'\ar d\ar p'\ar \gamma (p;p',k') \delta(\ar p\ar - \ar p'\ar
-\ar k'\ar)
\\ \nonumber
\left( f_p (1+f_{p'})(1+f_{k'})-f_{p'}f_{k'}(1+f_p)\right) \ ,
\ea
with
\be
\gamma(p;p'=p-k',k')= \left( \frac{1}{81\mu^4\cd 2 c_s^2} \right)
\ar  p\ar^2 \ar k'\ar^2 \ar p' \ar^2
\ .
\ee

We can analogously reduce the second term of Eq.~(\ref{collisionterm}),
that amounts essentially to the exchange $p\to p'$ with respect to the $C^{1 \to 2}_a$
piece,   as the splitting function is totally symmetric in its three arguments.
Thus, we won't write here the analogous equation for $C^{1\to 2}_b$.

We linearize the collision term in the first order in the deviation from equilibrium,
which we parametrize as
\be
f^1_p= \frac{X(x)}{T} f^0_{p} (1+f^0_{p}) \frac{\chi_p}{T} \ .
\ee

The last line in Eq.~(\ref{coll-term}) containing the distribution functions becomes
\be
 \label{fstructure}
F(p;p',k') \equiv \frac{X(x)}{T}
\frac{\chi_{p'} -\chi_{k'}-\chi_p}{T} f^0_{p} (1+f^0_{p'})(1+f^0_{k'}) \ .
\ee

We then expand the deviation from equilibrium in terms of the function basis
$\chi(p)/T= \sum_n \chi_n \psi_n(p/T)$. Then the projected, linearized collision
operator turns out to be
\ba \label{collisiontogether}
\la \psi_m \ar  \frac{\delta C[f]}{\delta f} \ar \psi_n\ra = \\
\nonumber
4\pi^2 \int \int \int_0^\infty \frac{dpdkdp'}{(2\pi T)^3} \psi_m (p)
\left( \delta(p'-p-k')
\gamma(p';p,k) F_n(p';p,k') \frac{X(x)}{T} - \frac{1}{2}(p\to p')\right)
\ea
where we define $F_n$ as the value of $F$ evaluated at $\psi_n$, instead of at  $\chi/T$.

To obtain the parametric dependence of the bulk viscosity  we define the dimensionless
quantities
\ba
\bar{\gamma} = \frac{\mu^4}{T^6} \gamma \\ \nonumber
\delta(p'-p-k')= \frac{1}{T} \delta(\frac{p'-p-k'}{T})
\ea
in terms of which, and extracting the factor $X(x)/T$,  the right hand
side of Eq.~(\ref{collisiontogether}) turns into
\be
\frac{X(x)}{T} \frac{T^5}{\mu^4} \la \psi_m \ar \frac{\bar{\delta
C}}{\delta f} \ar \psi_n \ra
\ee
where the energy-dimension is explicit, since $C_{mn}$ is now a function
of the ratios of momenta over temperature alone.
One can then check the dimension of the projected Boltzmann equation
\be
\int \frac{d^3p}{(2\pi T)^3} S(p) = -\frac{T^6}{\mu^4} \frac{\bar{\delta
C}(\chi/T)}{\delta f}
\ee
that matches the defining Eq.~(\ref{defofs}) above.

Finally, the parametric dependence of the viscosity, following from
Eq.~(\ref{bulkvisco}), is given as
\be \label{bulkresult}
\zeta = \frac{m_s^4}{T} \la \bar{S} \ar  \left(\frac{\bar{\delta
C}}{\delta f}\right)^{-1} \ar \bar{S} \ra \ .
\ee

\subsection{Numerical evaluation}
\label{Sec-num}

Once the parametric dependence of the bulk viscosity is known, all that remains is to evaluate a
numerical factor. A subtle point comes  in choosing an appropriate trial function family
$\psi_i$, so that all integrals converge appropriately in both their infrared (IR) and ultraviolet
(UV) domains.
For example, the natural family of orthonormal functions in the
interval $(0,\infty)$, the Laguerre functions, would yield
\be
\psi^m(p) = \frac{\sqrt{2}\pi e^{-p/2}}{p} L^m(p)
\ee
with conventional $L^m$ Laguerre polynomials.
The $1/p^2$ from two such functions, together with a $1/p$ from the
Bose-Einstein factor $(e^p-1)^{-1}\to p^{-1}$ and our splitting function
$\gamma \propto p^2$, make the integral in Eq.~(\ref{collisiontogether}) infrared
divergent. Fortunately, it is not necessary to use an orthonormal function family
\footnote{We thank Guy Moore for this observation}, as long as one is not
interested in the function $\chi$ itself, but only on its projection to
obtain the transport coefficient.  This can be shown with minimum
linear algebra by changing basis from an orthonormal  to an
arbitrary $\psi_i$ family function. We therefore choose
\be
\psi_m(p/T) = \frac{(p/T)^m}{(1+(p/T)^{m+5})} \ .
\ee

One should also notice that there is a zero-mode of the collision
integral, visible in the last line of Eq.~(\ref{fstructure}), that
vanishes when $\psi=p$ for $p=p' +k' $. The collision operator is not
invertible, but as observed in \cite{Arnoldbulk}, this does not suppose a
problem as one can add to $C$ an arbitrary constant $\lambda$ times the
projector over this zero mode
$$
C \to C + \lambda \ar E_p p^2 \ra \la E_p p^2 \ar
$$
since its projection over the source in Eq.~(\ref{bulkresult}) vanishes.
Note a factor of $f^0(1+f^0)$ can be multiplied to the vector $\ar E_p p^2
\ra$ without numerically affecting the result, as in Ref.
\cite{Arnoldbulk}.

\begin{figure}
\hspace{4cm}
\psfig{figure=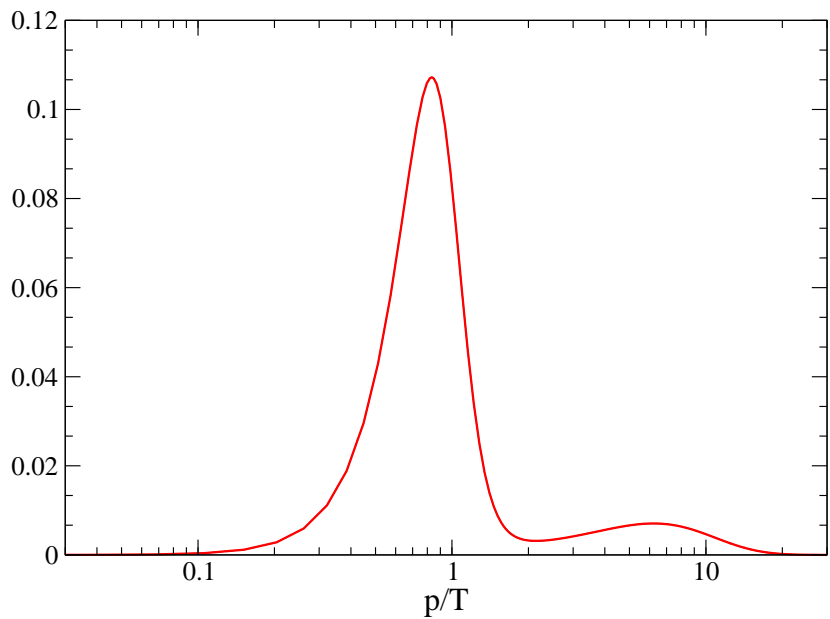,height=4.0in,angle=-90}
\caption{ Integrand for the last integral in Eq. (\ref{collisiontogether})
yielding the dimensionless $\bar{\zeta}$,
showing IR and UV integrability.}
\label{fig:conv}
\end{figure}

In table \ref{tablita_conv} we show the fast convergence with the number
of functions employed (size of the linear system).
\begin{table}
\caption{Convergence of the pre-coefficient of $\zeta$ with the
number of functions. First column: number of functions in the
family. Second column: result with zero-mode subtraction based on
the function $E_p p^2$. Third column: result with zero-mode
subtraction based on the function $E_p p^2 f_p (1+f_p)$. This
pre-coefficient can be thought of as $\bar{\zeta}=\zeta(m_s=1,T=1)=
\frac{T}{m_s^4} \zeta$. }{\label{tablita_conv}} \hspace{4cm}
\begin{tabular}{|ccc|}\hline
$m$ & $\bar{\zeta_1}$ & $\bar{\zeta_2}$ \\
1 & 0.00978 & 0.0093 \\
3 & 0.0109  & 0.01057\\
5 & 0.0110  & 0.01067\\
7 & 0.0110  & 0.01070\\
\hline
\end{tabular}
\end{table}
The integration is performed with a Gaussian grid,
there is no sensitivity to UV or IR cutoffs, as shown in
Fig.~\ref{fig:conv}, nor to the parameter $\lambda$
that fixes the zero-mode subtraction. In the table however we give two
sets of numbers showing that the result is essentially equivalent should
the factor $f_0(1+f_0)$ be omitted in the zero-mode subtraction.

\section{Discussion}
\label{Sec-disc}

We have established that the bulk viscosity coefficient associated to 
the normal fluid component of a 
cold CFL superfluid is given by
\be
\label{finalvisco} \zeta_{CFL} = 0.011 \frac{m_s^4}{T}
\ee
at
first order in the conformal-breaking parameter $m_s^2/\mu^2$.
This is a remarkable result for several reasons. First, it is much
smaller than the shear viscosity already reported in Ref.
\cite{Manuel:2004iv}. The reason for being so is that the two
coefficients are governed by different sort of processes
(collinear splitting for bulk viscosity, large-angle collisions
for shear viscosity), which occur at different rates. Second, at
the order we computed it is seen to be independent of the chemical
potential and the superconducting gap $\Delta$.
 Third, due to the dynamics being dominated by the superfluid phonon,
which is a  Goldstone  boson which remains always massless,
it is not exponentially suppressed as might be thought
of based on a calculation involving gapped degrees of freedom,
as, for example, those due to quarks \cite{Madsen:1999ci} or to kaons \cite{Alford:2007rw}.
Thus, it is the leading contribution at very small temperatures $T \ll m,\Delta$, where
$m$ is the energy gap associated to the lightest massive mode of the CFL phase. This is the
temperature regime that we call ``cold''.

The result in Eq.~(\ref{finalvisco}) could have been anticipated from
the mean free path for small angle collisions discussed in our previous
work \cite{Manuel:2004iv}, namely
\be
\lambda_{\rm small} \propto \frac{\mu^4}{T^5} \ .
\ee
A quick estimate of the parametric dependence of the bulk viscosity
would be
\be
\zeta \propto  \lambda_{\rm small} \times n \times \la p \ra \times
{\mathcal{C}}^2 \propto \frac{m_s^4}{T}
\ee
in terms of the phonon number density $n\propto T^3$ and average
momentum $\la p \ra \propto T$, and of the conformal breaking parameter
${\mathcal{C}}= \left( \frac 13 - c_s^2 \right) \sim
m_s^2/\mu^2$. This immediately yields Eq.~(\ref{finalvisco})
up to the numerical factor.
Note that the shear viscosity however has a different parametric behavior
with the temperature, as small angle collisions are very inefficient for
transferring transverse momentum. This is seen in the calculation in
\cite{Manuel:2004iv} by the appearance of near-zero modes that appear in
the computation of the $2\to 2$ collision operator and make collinear
splitting irrelevant there \footnote{The \emph{exact} zero mode appearing in
Eq.~(\ref{fstructure}) however, is not a separation from equilibrium
as it maintains the detailed balance relation $C[f_0+{\rm zero}]=0$ and
can be subtracted. This explains the failure of relations one could
have guessed such as $\zeta \propto m_s^4\eta/\mu^4$, that seem to fail
analogously in $\phi^4$ theory \cite{Jeon:1994if} but hold in the
quark-gluon plasma \cite{Arnoldbulk}.}.

While our bulk viscosity result diverges in the limit $T \to 0$, it should be kept in
mind \cite{Manuel:2004iv} that when the temperature diminishes the mean free path of the
phonon becomes large, and at some point the  hydrodynamical description of the phonon fluid is meaningless
(one rather has free streaming of phonons).
Then only the perfect superfluid with no dissipation remains. For astrophysical applications, and
assuming that the radius of the compact star is of the order of $R \sim 10$ km, this happens at
$T \sim 0.06$ MeV \cite{Manuel:2004iv}.

Although we have computed the bulk viscosity in a fluid at rest,
it is possible to extract from our results the frequency dependent
bulk viscosity needed for astrophysical applications. Based on our results, a quick estimate of a so defined
frequency-dependent bulk viscosity has recently appeared \cite{Alfordslides}.
The results are very interesting and suggest that the rate of equilibration of bulk distortions in a
hypothetical quark star, at physically relevant frequencies, receive
contributions not only from weak-equilibration processes but also
from the phonon-splitting processes that we have studied here
\cite{Alfordslides}.
Although quark-gluon equilibration times are short, the phonon-system
is described by the weakly coupled effective Lagrangian of Son, and
the bulk viscosity is proportional to the (small) scale-violating
parameter. This corrects one's first intuition about neglecting
strong interaction phenomenology altogether in the belief that strong
interactions should permanently be in equilibrium.

There is also another subtle point. In the existing literature where one needs a bulk viscosity
coefficient to perform the analysis of the fate of the r-modes, the
computation is performed by analyzing
the energy dissipated after a compression or rarefaction over one period, $\tau = 2 \pi/\omega$,
where $\omega$ is the frequency of the fluctuation. Thus
\be
\label{dissip}
\langle \frac{dE_{\rm dis}}{dt} \rangle = \frac{\zeta}{\tau} \int^\tau_0 dt \left( \nabla \cdot {\bf V} \right)^2
\ ,
\ee
where ${\bf V}$ is the hydrodynamical velocity.  The resulting bulk viscosity coefficient is then given
also as a function of $\omega$. While this relation is valid for a normal fluid, it should be generalized
for a relativistic superfluid. In a superfluid there are at least three bulk viscosities, and unless
the remaining coefficients vanish for  CFL quark matter, they should contribute to dissipation in an oscillatory
compression or rarefaction of the system, and should contribute to the right hand side of Eq.~(\ref{dissip}).

It is thus urgent a computation of the remaining viscosities of the CFL superfluid, as well as
a careful study of its low temperature hydrodynamics. Both are required for the study of the
r-modes of a hypothetical compact star made of CFL quark matter. Let us point out that
the relevance of the existence of several viscosities in
a superfluid neutron star has only been emphasized in  very recent publications
\cite{Andersson:2005pf,Andersson:2006nr,Gusakov:2007px}.
In particular, only in Ref.~\cite{Gusakov:2007px},
 all the bulk viscosity coefficients in a neutron superfluid have been computed.

\ack
We thank M. Alford, A. Dobado, L. Garay, M. Mannarelli, G. Moore and  D. T. Son
for useful discussions and remarks.
 Our work has been supporte by the grants
FPA 2004-02602, 2005-02327, PR27/05-13955-BSCH and AYA 2005-08013-C03-02.

\end{document}